# Dual Lepton Mass-Degeneracy-Deviation Quantities, Quasi-Degenerate Neutrinos and the Low Energy Fine Structure Constant

E. M. Lipmanov

40 Wallingford Road # 272, Brighton MA 02135, USA

## Abstract

Exponential lepton flavor mass-ratio phenomenology with an extreme value problem for the neutrinos predicts small Quasi-Degenerate (QD) neutrino masses, $m_\nu \cong (0.14-0.30)$ eV at 99% CL, and small solar-atmospheric oscillation hierarchy parameter $r \cong \Delta m^2_{sol}/\Delta m^2_{atm} \ll 1$, compatible with neutrino mass and oscillation data. An attempt is made to discover the physical meaning of the exponential $\exp(\pm 5)$ which emerges as an empirical parameter in that description. Accurate relations between the data value of the fine structure constant and integer 5 are observed starting with a bare value $\alpha_o = \exp(-5)$. The value $\alpha_o$ as a physical constant unifies dimensionless low energy gauge and Higgs-scalar coupling constants of the lepton electroweak-theory-sector and QD-neutrino mass ratios in conformity with duality and hierarchy rules in the discussed phenomenology of lepton flavor physics.

## 1. Introduction

The type of the neutrino mass pattern and the physical meaning of the empirically small neutrino oscillation solar-atmospheric hierarchy parameter are basic problems in neutrino physics. The neutrino mass type problem may be solved by the coming precise neutrino mass and oscillation experiments. New ideas in favor of a definite type of neutrino mass pattern may be appropriate and stimulating. In ref.[1] a dual relation between the neutrino and Charged Lepton (CL) mass ratios $x_n$, more accurately — $(x_n-1)$, is suggested, which is in favor of the widely discussed in the literature QD-neutrino mass type, determines the absolute QD-neutrino masses through the oscillation mass-squared differences and points to a connection between the solar-atmospheric hierarchy parameter $\Delta m^2_{sol}/\Delta m^2_{atm}$ and basic physics.

In this paper, new lepton flavor mass-ratio phenomenology is considered. It is assumed that the true, finite electroweak corrections to the bare lepton mass ratios, or lowest order mass-ratio values, are of several percent and can be meaningfully neglected in that consideration. The discussion involves lepton mass ratios, mass-ratio hierarchies, the special Mass-Degeneracy-Deviation (MDD) $(x_n-1)$-quantities and low energy electroweak coupling constants. No mixing considered between lepton mass eigenstates. A bare value $\alpha_o$ of the fine structure constant is pointed out which is smaller than the data value $\alpha_{Data}$ by about 8%. A sequence of empirical relations of substantially growing accuracy is observed between the data value of the fine structure constant $\alpha_{Data}$ at zero momentum transfer and the bare value $\alpha_o$. It is argued that this bare value $\alpha_o$ is a primary physical constant in lepton flavor physics beyond the Standard Model. The discussed lepton flavor phenomenology is based on *duality and hierarchy rules*: dual neutrino and CL MDD-quantities are connected by equal nonlinear hierarchy-relations of these quantities.

In Sec.2, main results of the lepton mass-ratio phenomenology with QD-neutrinos [1,2] are reviewed and augmented by new emphasizes and focus on the many-sided solar-atmospheric hierarchy parameter *r* as a link between neutrino oscillation data and basic physics. In Sec.3, accurate empirical relations between the fine structure constant $\alpha$ at $Q^2=0$ (data value) and exponential exp(-5) are observed, in accord with lepton mass-ratio indications. Sec.4 contains a discussion and conclusions. In Appendix, the small QD-neutrino mass scale $m_\nu$ is tentatively described by a drastic mass-ratio hierarchy of four lepton mass levels: $m_\tau$, $m_\mu$, $m_e$ and $m_\nu$.

## 2. Extreme value problem for the QD-neutrino mass scale and the solar-atmospheric hierarchy parameter

The idea that the CL and neutrino MDD ($x_n-1$)-quantities are in a dual relation to each other is described in [2] by two dual solutions with large and small exponents of the same nonlinear equation:

$$(x_2{}^k - 1)^2 = (\xi^k)(x_1{}^k - 1), \quad k \geq 1, \tag{1}$$

$x_n$ are the lepton mass-ratios, $x_n \to X_n = m_{n+1}/m_n$ for the CL with $m_1 = m_e$, $m_2 = m_\mu$ and $m_3 = m_\tau$ at a common low scale[1], or neutrino mass ratios $x_n$; k is an arbitrary value of the mass-ratio powers.

Equation (1) is an extension of the CL empirical mass-ratio hierarchy equation[2]

$$(m_\tau/m_\mu)^2 = \xi(m_\mu/m_e), \quad \xi_{exp} \cong 1.37. \tag{2}$$

The physical meaning of Eq.(1) is a *symmetry* of the nonlinear MDD-hierarchy, described by relation (1), under the interchange of CL and neutrinos — a presumable generic hierarchy rule in lepton flavor physics, and an intimate quantitative uniting condition for the two empirically very different lepton mass spectra of the neutrinos and CL beyond the electroweak theory.

---

[1] The difference between running and physical masses of the CL can be disregarded at the considered accuracy to within a few percent [3].

[2] The empirical coefficient $\xi$ in (2) can be represented as an expansion in terms of the small low energy semi-weak coupling constant $\alpha_W \cong 5\exp(-5)$, see (21) below, $\xi \cong \sqrt{2}\,(1 - \alpha_W) \cong 1.367$.

The stated exponential solutions of equation (1) with large and small exponents are:

$$X_1 \equiv m_\mu / m_e \cong \xi \exp\chi, \ X_2 \equiv m_\tau / m_\mu \cong \xi \exp(\chi/2), \ \chi_{exp} \cong 5, \tag{3}$$

$$x_2 \equiv m_3 / m_2 = \exp(a_k r), \ x_1 \equiv m_2 / m_1 = \exp(a_k r^2), \tag{4}$$

$$a_k r << 1, \ r_{exp} \cong (\Delta m^2_{sol} / \Delta m^2_{atm}),$$

for the CL and neutrino[3] mass ratios respectively (first powers).

CL solution (3) is independent of the power k in Eq.(1), but the solution for the neutrino mass ratios is not: the coefficient $a_k$ in (4) is a function of k,

$$a_k = \xi^k / k. \tag{5}$$

The interesting point here is that the coefficient $a_k$ in (5) as a function of k has a unique, absolute minimum [2]:

$$a_{min} = a_{ko} = e \ \log\xi \cong 0.85 \tag{6}$$

at $k = k_o \equiv 1/\log\xi$, where $e$ is the base of natural logarithms[4].

At $k=k_o$, Eq.(1) is changed to

$$(x_2{}^{ko} - 1)^2 = e(x_1{}^{ko} - 1), \quad k_o \cong 3.2. \tag{7}$$

In the present phenomenology, there are two types of physical quantities related to the neutrino mass spectrum: independent of the coefficient $a_k$ quantities, and other ones which depend on $a_k$.

It seems reasonable to assume that the physical characteristics of the actual neutrino mass spectrum should be independent of the arbitrary value k in (1), as it is in the case of CL, and are related to the solution of the extreme value problem for the coefficient $a_k$: the actual generic lepton hierarchy equation is given by (7) instead of (1), and the actual MDD-values of the neutrino mass spectrum are the minimal possible values given by

$$(x_2-1)_{min} \cong a_{min} \; r, \quad (x_1-1)_{min} \cong a_{min} \; r^2, \quad a_{min} \cong 0.85. \qquad (8)$$

[3] A sequence of neutrino masses $m_1 < m_2 < m_3$ is chosen. An alternative solution with $x_2$ and $x_1$ interchanged in (4) is also possible with the same conclusions.

[4] The value $a_{min}$ in (6) can be expressed through the semiweak interaction coupling constant $\alpha_W$, see footnote [2], $a_{min} \cong e(\log\sqrt{2} - \alpha_W) \cong 0.85$.

The QD-neutrino solar-atmospheric hierarchy parameter

$$r \cong (x_1^2-1)/(x_2^2-1) \cong (m_2-m_1)/(m_3-m_2) \qquad (9)$$

is independent of $a_k$ in contrast to the relative, dimensionless-made neutrino mass-squared differences $(x_1^2-1) \cong 2a_k r^2$ and $(x_2^2-1) \cong 2a_k r$. With parameter $r$ independent of k, the minimum value of the coefficient $a_k$ in (8) means minimal possible values of both the neutrino MDD-quantities and neutrino mass ratios, as a physical condition for the actual QD-neutrino mass pattern.

The two exponential solutions of Eq.(7) with large and small exponents for the CL and QD-neutrino mass ratios are given by

$$X_1 \cong \xi \exp 5, \quad X_2 \cong \xi \exp 5/2, \qquad (10)$$

$$x_2 \cong \exp(0.85r), \quad x_1 \cong \exp(0.85r^2), \quad 0.85r_{QD} \ll 1. \qquad (11)$$

Since the neutrino mass-ratio coefficient $a_{ko}$ is fixed in (6) and is not small, the neutrino solution (11) leads to the definite prediction of a small solar-atmospheric hierarchy parameter in the QD-neutrino scenario:

$$r_{QD} \ll 1, \qquad (12)$$

in agreement with neutrino oscillation data: $r_{exp} \ll 1$. So, the supposition that the solution of Eq.(7) with small exponents (11) describes the observable physical QD-neutrino mass ratios is supported by neutrino oscillation data (it passed the first necessary experimental test).

It should be noticed that the solutions (10) and (11) for the mass ratios of the CL $X_{1,2}$ and QD-neutrinos $x_{2,1}$ are very different in form and value. On the contrary, the corresponding solutions for the MDD-quantities of the CL [$(X_2 - 1) \cong \xi$ (exp5/2), $(X_1 - 1) \cong \xi$ (exp5/2)$^2$] and QD-neutrinos [$(x_2 - 1) \cong 0.85 \, r$, $(x_1 - 1) \cong 0.85 \, r^2$] are respectively large and small and fully conformable (same hierarchical patterns).

Finally, the QD neutrino mass scale, pertaining to the *minimal* neutrino MDD-quantities, is given by

$$m_\nu \cong (\Delta m^2_{atm}/1.7r)^{1/2} = (\Delta m^2_{sol}/1.7r^2)^{1/2} = \Delta m^2_{atm}/(1.7\Delta m^2_{sol})^{1/2}. \quad (13)$$

With the best-fit values of the atmospheric and solar mass-squared differences and solar-atmospheric hierarchy parameter[5] from [4],

$$\Delta m^2_{atm} = 2.2 \text{x} 10^{-3} \text{ eV}^2, \ \Delta m^2_{sol} = 8.1 \text{x} 10^{-5} \text{ eV}^2, \ r_{bf} = 0.035, \quad (14)$$

the neutrino mass scale is given by

$$m_\nu \cong 0.19 \text{ eV} \quad (15)$$

to within a few percent.

With the 99% CL range [5] for mass-squared differences:

$$\Delta m^2_{atm} = (1.7 \div 3.3) \text{x} 10^{-3} \text{ eV}^2, \ \Delta m^2_{sol} = (7.2 \div 8.9) \text{x} 10^{-5} \text{ eV}^2, \quad (16)$$

the neutrino mass scale is

$$m_\nu \cong (0.14 - 0.30) \text{ eV}. \quad (17)$$

These neutrino mass values are compatible with the upper limits from known data, see [4-6].

All important physical characteristics of the neutrino mass spectrum, (8) and (11)-(13), are expressed through one parameter *r* which has a special, many-sided physical meaning in the QD neutrino scenario:

1) It represents the solar-atmospheric hierarchy parameter

---

[5] The estimation of the global best-fit value of the solar-atmospheric hierarchy parameter in [4] is $r_{bf} = 0.035$. By an obvious reason, the notation "$\alpha$" for the solar-atmospheric hierarchy parameter ($\Delta m^2_{sol}/\Delta m^2_{atm}$) in [4] cannot be used here (see (20) and Sec.3).

$$r_{\exp} = (\Delta m^2_{sol}/\Delta m^2_{atm}).$$ (18)

2) The known large hierarchy of the CL mass ratios predicts small value of the neutrino oscillation hierarchy parameter $r$ by the dual relation between neutrino and CL MDD-patterns (small versus large mass-ratio exponents), (10) and (11). So, the positive result of neutrino oscillation experiments $r_{\exp} \ll 1$ is strongly in favor of QD neutrinos.

3) The parameter $r$ is a naturally small[6] phenomenological factor in the QD neutrino mass-ratio exponents (11).

4) It is independent of the neutrino mass scale.

5) It measures the hierarchy of the deviations from neutrino mass degeneracy $r \cong (x_1^2-1)/(x_2^2-1) \ll 1$.

There is an analogy between *the only two empirically known lepton MDD hierarchies* — of the QD neutrino MDD-quantities (parameter $r$) and of the MDD-quantities of the CL (parameter R):

$$R \equiv (X_2^2-1)/(X_1^2-1) \cong (m_\tau/m_\mu)^2/(m_\mu/m_e)^2 \cong \exp(-5) \ll 1.$$ (19)

These hierarchies are probably related quantitatively, see below.

6) The condition $r \neq 0$ in (11) determines the violation of the QD neutrino mass eigenstate symmetry by small, hierarchical neutrino mass splitting - it hints at some dynamical connection between the solar-atmospheric hierarchy parameter $r$ and the neutrino mass splitting interaction.

7) By the known neutrino oscillation data, the value of the hierarchy parameter $r$ may be close to the value of the semiweak analogue of the low energy fine structure constant [1],

$$r = \lambda \alpha_W = \lambda g_W^2/4\pi \cong 0.034 \lambda \cong \lambda/30,$$ (20)

---

[6] As it is a measure of the deviation from mass eigenstate symmetry of QD-neutrinos, and Ref.[7].

$\lambda$ is a numerical factor of order 1, probably in the range $\lambda \cong$ $(1 \pm 0.2)$, compatible with the estimation of the best-fit data value [4]: $r_{bf} = 0.035$.

With the conditions above, the neutrino oscillation hierarchy parameter $r$ (18) should have a more basic physical meaning than the solar and atmospheric neutrino oscillation mass-squared differences have separately.

8) There is an interesting, *relevant* quantitative coincidence between the low energy dimensionless semiweak coupling constant $\alpha_W$ and the integer 5:

$$\alpha_W \cong 5 \exp(-5) \cong 0.034. \qquad (21)$$

With $r = \lambda \alpha_W$, this coincidence points to a relation between the exponents $r$ and $\chi$ in (10) and (11) respectively,

$$r \cong 5\lambda \exp(-5), \quad r \cong \lambda \chi \exp(-\chi). \qquad (22)$$

Relation (22) is a quantitative connection between the small and large exponents of the lepton mass ratios in the solutions (3) and (4), or (10) and (11). This relation is an interesting inference from the conditions 6), 7) and 8).

The lepton mass-ratio solutions (10) and (11) are intertwined by the parameter $\xi$. The known very large deviation from the CL mass-degeneracy and a possible weakly broken QD neutrino mass symmetry are two solutions of the generic Eq.(7). Such a relation between two MDD solutions of the same equation is termed "dual" relation in the present discussion. It is a quantitative relation between the hierarchically divergent CL mass pattern and hierarchically convergent mass pattern of the QD-neutrinos[7].

---

[7] There is an empirical similarity between the mass-ratio patterns of the Dirac particles — quarks and CL. If the QD-neutrinos are Majorana particles, the neutrino-CL duality pointed out above may suggest a more general Dirac-Majorana

The dual relation between the neutrino and CL dimensionless MDD-quantities is substantiated by connection (22) between the small and large exponential factors $r$ and $\chi$:

$$[(m_3/m_2)^2-1] \cong 1.7 \; \lambda \; \chi \; \xi^2 \; /[(m_\tau/m_\mu)^2-1],$$

$$[(m_2/m_1)^2-1] \cong 1.7 \; \lambda^2 \chi^2 \; \xi^2 \; /[(m_\mu/m_e)^2-1]. \qquad (23)$$

These connected neutrino and CL MDD-quantities obey the generic hierarchy equation (7). The large CL mass ratios $m_\tau^2/m_\mu^2$ and $m_\mu^2/m_e^2$ have direct physical meaning of MDD-quantities for the CL, while the inverse small CL mass ratios $(m_\mu^2/m_\tau^2)$ and $(m_e^2/m_\mu^2)$ have a dual, indirect MDD-meaning: they are proportional to the small MDD-quantities of the QD-neutrinos by relations (23).

From (23) and (19), it follows

$$r_{QD} \cong (x_1^2-1)/(x_2^2-1) \cong 5\lambda(m_\tau/m_\mu)^2 /(m_\mu/m_e)^2 \equiv 5\lambda \; R. \qquad (24)$$

Observe that the only two empirically known lepton MDD-hierarchies are related to one another in the QD neutrino scenario: the ratio of these hierarchies in (24) is $r/R \cong 5\lambda$. With that, the well known very large CL mass-ratio squared hierarchy $R \cong 1/151$ from (19) leads once more to a quantitative prediction of a large hierarchy of the solar and atmospheric neutrino oscillation mass-squared differences, $r \cong \lambda/30$.

In summary, the generic nonlinear lepton MDD hierarchy equation (7) with dual neutrino and CL mass ratio solutions (10) and (11) predicts QD neutrinos, small absolute neutrino masses (17) and small solar-atmospheric hierarchy parameter $r$, in agreement with neutrino mass and oscillation data [4-6].

A special feature of the QD neutrino pattern, as distinct from the hierarchical and inverted patterns and from the now

---

MDD-duality pattern as a rule in flavor physics. The known divergence of mass spectra of both the CL and quarks should be related to the mass Quasi-Degeneracy of the neutrinos, as a basic relation in flavor physics.

obsolete case of massless neutrinos, is a clear regularity in the lepton flavor physics: a dual relation between CL and neutrino MDD $(x_n-1)$-quantities, which is described by two exponential solutions of Eq.(7) with large and small exponents $\chi$ and $r$ probably connected by the relation (22).

Empirical evidences are pointing to an emergence of the exponential exp(±5) as a common unifying physical constant: on the one hand, the CL mass ratios are expressed through the exponential $e^5$ by solution (10) at the dominant level, with the necessary electroweak corrections to be at a subdominant level; on the other hand, with relation (22), the QD-neutrino mass ratios are likely determined at the dominant level by the low energy semiweak dimensionless coupling squared $\alpha_W$, also related to that constant. With (20)-(22), the solar-atmospheric hierarchy parameter $r$ is a probable link between the neutrino oscillation data and new flavor physics.

Integer 5, as a likely discrete symmetry characteristic of the QD neutrino mass matrix with approximate bimaximal mixing, is discussed in Ref.[8].

The possibility of baryogenesis with QD-neutrinos, $m_v \geq 0.1$ eV, is considered in Ref.[9].

A discussion of QD-neutrino masses within a theoretical seesaw-II framework is recently considered in Ref.[10].

### 3. The fine structure constant at zero momentum transfer and exponential $e^{-5}$

Why the exponential $e^{\pm 5}$? There are at least two possible answers to this unavoidable question: 1). The empirical appearance of the exponential exp(±5) in the CL mass ratios (10), in the low energy semi-weak coupling constant (21), in the MDD-hierarchies (19) and (22) and neutrino mass ratios (11) is a set of coherent coincidences. 2). The appearance of the exponential $e^{\pm 5}$ is related to new physics where it plays a universal dimensionless physical constant. In the interesting second case, the approximate relation between the low energy semiweak coupling constant $\alpha_W$ and that exponential in the relevant relation (21) should be not a mere coincidence. Further relations between the gauge coupling constants of the electroweak theory [11] and the exponential $e^{-5}$ should be expected. It is observed below that there are several explicit successive relations of growing accuracy between the *highly informative precision data value $\alpha_{Data}$ of the fine structure constant and the exponential $e^{-5}$.*

1) The approximate relation

$$\alpha \cong \alpha_o \equiv e^{-5}, \tag{25}$$

where $\alpha \cong 1/137$ is the empirical value [12] of the fine structure constant, is correct to within ~8%. Approximation (25) is prompted by the lepton mass-ratio indications and the relevant relation (21) for the semiweak constant $\alpha_W \cong 5\exp(-5)$ plus the approximate empirical relation [12] for the Weinberg mixing angle $\theta_W$ of the electroweak theory: $\sin^2\theta_W \cong 0.2$.

Relation (25) is a remarkable one. With (10)-(22), and the electroweak theory [11] relation $\alpha_W = \alpha/\sin^2\theta_W$, it follows

$$\alpha \cong \alpha_o, \ \sin^2\theta_W \cong -1/\log\alpha_o = 0.2, \ \alpha_W \cong -\alpha_o\log\alpha_o, \tag{26}$$

$$m_\mu/m_e \cong \sqrt{2}/\alpha_o, \ m_\tau/m_\mu \cong \sqrt{(2/\alpha_o)}, \ R \cong \alpha_o, \tag{27}$$

$$(m_3/m_2-1)_{nu} \cong 0.85r, \ (m_2/m_1-1)_{nu} \cong 0.85\,r^2, \ r \cong \lambda\,\alpha_W, \tag{28}$$

where $r$ and R are the MDD-hierarchies of the neutrino and CL masses (9), (19) and (22), $\lambda \cong 1$.

Relations (26)-(28) are correct to within <~8%. The "bare" value of the fine structure constant $\alpha_o$ unifies the low energy electroweak gauge coupling constants and lepton mass-ratios[8]; five independent dimensionless coupling constants of the low energy electroweak lepton interactions[9], with the electromagnetic field and $W^\pm$-field (gauge interactions) and scalar Higgs-field interactions ($f_e$, $f_\mu$ and $f_\tau$ — coupling constants with the scalar field of the electron, muon and tauon respectively)[10],

$$f_\mu = (\sqrt{2}/\alpha_o)\, f_e, \quad f_\tau = \sqrt{(2/\alpha_o)}\, f_\mu, \tag{29}$$

are unified via the parameter $\alpha_o$; one coupling constant with the scalar field is left free (e.g. $f_e$, i.e. electron mass $m_e$).

2) Observe now, as a *factual* condition with $(\alpha^{-1})_{Data} \cong 137.036$ from [12], that the following extension of the approximate relation (25):

$$\exp 2\alpha\, \log(\exp\alpha\, /\alpha) \cong \log(1/\alpha_o) \tag{30}$$

is correct to within $\sim 3 \times 10^{-6}$. This large increase in accuracy is achieved not by consecutive adjusting terms, but by the addition to the relation (25) of two close to unity and clearly related exponential factors ($\exp\alpha$) and $(\exp\alpha)^2$ without fine tuning[11].

---

[8] With the estimation of neutrino mass scale $m_v$ in (17), the ratio of that scale to the electron mass $m_e$ should be $(m_v/m_e) = \eta\, \alpha_o^3$, where the coefficient $\eta$ is $\eta = (0.9 \div 1.9)$ at the 99% CL.

[9] The electroweak theory consistently describes all interactions of the leptons, with empirical values of the masses and gauge coupling constants (not included gravity).

[10] For the CL: $m_l = f_l \langle\varphi\rangle$, $l = e, \mu, \tau$, $\langle\varphi\rangle$ is the vacuum expectation value of the scalar field $\varphi$.

[11] Eq.(30) is a phenomenological equation, presumably derivable in a future finite nonperturbative theory of

If considering (30) as an equation for the unknown $\alpha$, the solution is

$$\alpha \cong 1/137.0383, \quad (\alpha - \alpha_{Data})/\alpha_{Data} \cong -1.7 \times 10^{-5}, \qquad (31)$$

compared to $[\alpha_o - \alpha_{Data}]/\alpha_{Data} \cong -0.08$.

So, the accurate relation (30) does support the choice of bare value $\alpha_o$ for the fine structure constant $\alpha$ as the proper one.

Another interesting feature of Eq.(30) is that after exponentiation,

$$(\exp\alpha /\alpha)^{\exp 2\alpha} \cong \exp 5, \qquad (30')$$

the difference between its right and left sides in (30') is equal to $(\alpha/\pi)$ to within $\sim 0.002$ — an impressive relative accuracy.

3) And so, a further extended relation

$$\exp 2\alpha \log(\exp\alpha /\alpha) \cong \log(1/\alpha_o - \alpha/\pi) \qquad (32)$$

between the integer 5 and the highly accurate data value of the fine structure constant [12],

$$(1/\alpha)_{Data} = 137.03599911(46), \qquad (33)$$

is satisfied to within $\sim 6 \times 10^{-9}$.

If considering (32) as an equation for the unknown $\alpha$, the solution is

$$\alpha \cong 1/137.0359948, \quad (\alpha - \alpha_{Data})/\alpha_{Data} \cong 3.1 \times 10^{-8}. \qquad (34)$$

It differs from the central data value of the fine structure constant (33) by about 10 S.D.

Note that the additional term $\alpha/\pi$ on the right side of (32) is like a "perturbation term" added to the "nonperturbative" basic relation (30).

The sequence of consecutive empirical relations of growing accuracy (25), (30) and (32) between the data value of the fine lepton masses and interactions if starting with the bare value $\alpha_o = \exp(-5)$ of the fine structure constant.

structure constant $\alpha_{Data}$ and the bare value $\alpha_o$ supports the independent indications in Sec.2 that the exponential $\alpha_o$ should be a universal physical constant in the phenomenology of lepton flavor physics.

4) Finally, a possible second "perturbative" term (e.g. $\alpha^2/4\pi$) raises the accuracy of relation (32) by about one order of magnitude, but transforms it into an highly accurate nonlinear equation for the unknown $\alpha$:

$$\exp2\alpha \ \log(\exp\alpha\,/\alpha) = \log(1/\alpha_o - \alpha/\pi + \alpha^2/4\pi), \qquad (35)$$

or, comp. (30'),

$$(\exp\alpha\,/\alpha)^{\mathbf{exp2\alpha}} + \alpha/\pi - \alpha^2/4\pi = \exp5. \qquad (35')$$

Indeed, the solution of Eq.(35) is given by

$$\alpha \cong 1/137.03599901, \ (\alpha-\alpha_{Data})/\alpha_{Data} \cong 0.7\times10^{-9}. \qquad (36)$$

This solution $\alpha$ of the phenomenological equation (35) agrees with the central data value of the fine structure constant at zero momentum transfer $\alpha_{Data}$ in (33) to within ~0.2 S.D.

Equation (35) is a *virtual accurate equation* for the fine structure constant $\alpha$ at zero momentum transfer with the exponential exp5 on the right side of Eq.(35') as the origin of the precise numerical solution (36).

So, the question of where does the specific numerical value of the fine structure constant at zero momentum transfer come from may have an answer. With the relations above, a unique connection between the fine structure constant $\alpha$ at $Q^2=0$ and the "bare" value $\alpha_o$ is possible, at least to within a few S.D. If the true numerical value of the fine structure constant at zero momentum transfer is reduced to integer 5 (alpha-genesis[12])

---

[12] The fine structure constant $\alpha$ at the particular momentum transfer $Q^2=0$ is a very special physical quantity because of its role in nonrelativistic quantum mechanics of bound

through the bare value $\alpha_o = e^{-5}$, all gauge coupling constants of the Standard Model might be expressed through the constant $\alpha_o$ by the renormalization group equations and Grand Unification [13].

## 4. Discussion and conclusions

In the discussed lepton mass-ratio phenomenology, two pairs $[(X_1-1),\ (X_2-1)]$ and $[(x_2-1),\ (x_1-1)]$ of MDD-values for the CL and neutrinos respectively are the primary physical quantities. They are appropriate and suggestive in the QD-neutrino scenario. With three lepton generations, there is one pair of large and hierarchical MDD-quantities for the CL, and one pair of small and hierarchical ones for the neutrinos. The mass ratios $x_n$ of the CL and QD-neutrinos are very different; in contrast, the values of the MDD-quantities $(X_n-1)$ and $(x_n-1)$ are large and small respectively but have analogous hierarchical patterns. Both lepton MDD-ratios (MDD-hierarchies) are empirically known physical quantities: the QD-neutrino MDD-hierarchy is nearly the neutrino oscillation hierarchy parameter $r \cong (\Delta m^2_{sol}/\Delta m^2_{atm}) \langle\langle 1$, while the CL MDD-hierarchy is nearly the ratio of squared mass-ratios $R \cong (m_\tau/m_\mu)^2 / (m_\mu/m_e)^2 \langle\langle 1$. They are related to one another by relation (24), $r/R \cong 5\lambda$, i.e. $r \cong \lambda/30$, $\lambda \approx 1$.

An extreme value problem for the neutrino $(x_n-1)$-quantities leads from Eq.(1), which includes an arbitrary parameter, to the basic hierarchy equation (7). It has two dual solutions with large and small exponents, which describe the CL and neutrino MDD-patterns and predict small QD neutrino masses, small solar-atmospheric hierarchy parameter $r_{QD} \ll 1$ and probably a

---

states enabling Life and Consciousness. The unique connection of $\alpha$ with integer 5 is a possible solution to the anthropic principle problem for the fine structure constant in our Universe.

quantitative connection between the two observable large MDD-hierarchies. The result of the neutrino oscillation experiments $r_{\exp} \ll 1$ is the first positive experimental test of QD neutrino mass pattern strongly predicted by solution (11). The small values, (13)—(17), of the QD neutrino mass scale is an interesting quantitative prediction: $m_\nu \cong (0.14 - 0.30)$ eV at the 99% CL.

In the QD-neutrino scenario, the parameter $r$ is a naturally small factor of the neutrino mass-ratio exponents with a many-sided physical meaning. It is a probable link between the neutrino oscillation data and new fundamental physics. With QD neutrinos, the small parameter $r$ determines the neutrino mass splitting by the solution (11), on the one hand; on the other hand, the empirical value of $r$ may be close to the low energy semiweak coupling constant $r = \lambda \, \alpha_W \cong 5\lambda \, \alpha_o$ with the coefficient $\lambda$ of order 1. This relation meets two conditions (a) a connection of the small hierarchy parameter $r$ (neutrino mass splitting) with the weak interaction dimensionless coupling constant $\alpha_W$ and (b) a connection between the exponential factors $r$ and $\chi$ of the QD-neutrino and CL mass ratios (11) and (10). The coming precision neutrino oscillation data will determine how close to unity is the value of the coefficient $\lambda$.

Accurate empirical relations between the fine structure constant at momentum transfer $Q^2=0$ and the integer 5 are observed, starting with the bare value[13] $\alpha_o = \exp(-5)$, which is prompted by indications from the lepton mass ratios (10)—(22) and the relevant, empirical relation $\alpha_W \cong 5 \, \alpha_o$. The physical constant $\alpha_o$ unifies the CL and QD-neutrino mass ratios, as well as the low energy lepton gauge and Higgs-scalar coupling constants in the electroweak model of leptons; it is the main common exponential factor in the discussed lepton MDD quantities in Sec.2 and gauge coupling constants in the lepton sector of the electroweak theory in Sec.3. As a result, the empirical functional dependence on integer 5 of the MDD $(x_n-1)$—quantities of the CL and QD-neutrinos resembles that of the low energy electroweak coupling constants in the succession $1/\alpha$ and $\alpha_W$ respectively, comp. (27) and (28).

---

[13] With the proper choice of the bare constant $\alpha_o$, the accurate empirical value of the fine structure constant (33) from the Particle Data [12] proved to be highly informative.

Conclusions.

1) As considered in Sec.2, a dual relation between two highly hierarchical pairs of MDD-quantities of the CL $[(X_1^2-1)_{CL} \cong (m_\mu/m_e)^2$, $(X_2^2-1)_{CL} \cong (m_\tau/m_\mu)^2]$ and neutrinos $[(x_2^2-1)_{nu} \cong 1.7\, r_{QD}$, $(x_1^2-1)_{nu} \cong 1.7\, r_{QD}^2]$, predicts small solar-atmospheric hierarchy parameter $r_{QD} \ll 1$ and small QD-neutrino masses[14] (13)-(17). The coming precision neutrino oscillation data will determine how close to unity is the coefficient $\lambda$ in the empirical relation $r = \lambda\, \alpha_W$.

2) An explicit bare value of the fine structure constant $\alpha_o = e^{-5}$ is pointed out in Sec.3. It uniquely determines the precise data value of the fine structure constant at zero momentum transfer by virtual equations (30)-(35). Probably, they suggest new fundamental physics enabling a connection between the gauge coupling constants of the electroweak theory and the lepton mass ratios through the exponential $\alpha_o = \exp(-5)$. It is argued that $\alpha_o$ may be a basic physical constant in new lepton flavor mass-ratio physics beyond the Standard Model.

---

[14] A test of QD neutrino mass type is in the sensitivity region of present experiments [14].

3) A *remarkable relation* is noted between the large and small lepton MDD-quantities and the low energy electroweak coupling constants $\alpha$ and $\alpha_W$: $(X_1^2-1)_{CL} \cong 2/\alpha_o^2$, $(X_2^2-1)_{CL} \cong (2/\alpha_o)$; $(x_2^2-1)_{nu} \cong 1.7\,\alpha_W$, $(x_1^2-1)_{nu} \cong 1.7\,\alpha_W^2$, $\alpha_W \cong 5\alpha_o$. So, the very small bare value $\alpha_o$ of the fine structure constant connects the hierarchically divergent CL masses with the hierarchically convergent Q-degenerate neutrino masses — another aspect of the stated dual relation between the MDD-quantities of the CL and neutrinos.

4) Both the CL and QD-neutrino MDD-ratios (9) and (19) are large MDD-hierarchies and *empirically known lepton physical quantities*, $1/R \cong [\,(m_\mu/m_e)\,/(m_\tau/m_\mu)\,]^2 >> 1$, $1/r \cong (x_2^2-1)_{nu}\,/(x_1^2-1)_{nu} \cong (\Delta m^2_{atm}/\Delta m^2_{sol}) >> 1$. The condition that the "nonlinear MDD-hierarchy relations" of the neutrinos and CL in (7) are equal:

$(x_2^{ko}-1)^2/(x_1^{ko}-1) \cong [\,(m_\tau\,/m_\mu)^2\,/\,(m_\mu\,/m_e)\,]^{ko} = e$, $k_o \cong 3.2$,      (37)

determines small absolute values (13)-(17) of QD-neutrino masses: $m_\nu \cong (0.14 - 0.30)\,eV$ at 99% CL.

The actual QD-neutrino MDD-quantities (and mass ratios) are determined by the extreme (minimum) value condition (8) through the neutrino oscillation hierarchy parameter $r$.

The many-sided solar-atmospheric hierarchy parameter $r$, primarily being the naturally small factor in the exponents of the neutrino mass-ratio solution (11), is a link between neutrino oscillation data and new lepton flavor mass-ratio physics.

5) *Duality and hierarchy rules in lepton flavor physics* beyond the electroweak theory:
(a) Dual relation between MDD-quantities of the CL and QD-neutrinos: they are solutions of the same equation (7) with large and small exponents $\chi$ and $r$ connected by relation (22) which determines the quantitative dual relations (23) between the CL and neutrino MDD-quantities.

(b) Large MDD-hierarchies of the neutrinos ($1/r \gg 1$) and CL ($1/R \gg 1$), and equal "nonlinear MDD-hierarchy relations" (37) of the CL and QD-neutrinos. The ratio of the QD-neutrino MDD-hierarchy ($r$) to the CL one (R) is close to the ratio of low energy fundamental electroweak physical constants $\alpha_W$ and $\alpha$:

$r \sim \alpha_W$, $R \sim \alpha$, $r/R \cong \lambda(\alpha_W/\alpha)$, $\lambda \approx 1$.

The statements (a) and (b) are two phenomenological rules in lepton flavor physics. They are not quite independent: neutrino-CL MDD-duality is a solution of the generic nonlinear lepton hierarchy equation (7). These empirical rules of lepton flavor physics are imparted by the two solutions (10) and (11) for the CL and QD-neutrino MDD-quantities in view of the lepton mass and neutrino oscillation data in Sec.2 plus the relation between these lepton MDD-quantities and the low energy basic electroweak quantities $1/\alpha$ and $\alpha_W$ discussed in Sec.3.

6) The conclusions listed above show the Quasi-Degenerate neutrino mass type to the best advantage[14].

I thank Valery Khoze and Lev Okun for reading the manuscript.

## References

[1] E. M. Lipmanov, Phys.Lett. **B567** (2003) 268.

[2] E. M. Lipmanov, hep-ph/0402124.

[3] H. Arason et al., Phys.Rev. **D46**, 3945, 1992.

[4] M. Maltoni, T. Schwetz, M. A. Tortola, J. W. F. Valle, New J.Phys. **6**, (2004) 122.

[5] A. Strumia, F. Vissari, hep-ph/0503245.

[6] J. N. Bahcall, M. C. Gonzalez-Garcia, C. Pena-Garay, JHEP 0408 (2004) 016; V. Barger, D. Marfatia, K. Whisnant, Int. J. Mod. Phys. **E12 (**2003) 569; A. Yu. Smirnov, Int. J. Mod.

Phys. **A19** (2004) 1180.

[7] J. A. Casas, J. R. Espinosa, A. Ibarra, I. Navarro, Nucl.Phys., **B556** (1999) 3.

[8] E.Ma, hep-ph/0409288; E.Ma, hep-ph/0409075 and references therein.

[9] H. Davoudiasl et al., Phys.Rev.Lett. **93**, 201301, 2004.

[10] R. N. Mohapatra, hep-ph/0412379.

[11] S. L. Glashow, Nucl. Phys. **22**, 579, 1961; S. Weinberg, Phys.Rev.Lett. **19**, 1264, 1967; A. Salam, In: Elementary Particle Theory, Almquist and Wiksell, Stockholm, p.367, 1968.

[12] Particle Data Collaboration, S. Eidelman et al.,Phys. Lett. **B592**, 1, 2004.

[13] Originally in SU(5) by H. Georgi, S. L. Glashow, Phys.Rev. Lett. **32**, 438, 1974; H. Georgi, H. R. Quinn, S. Weinberg, Phys.Rev.Lett. **33**, 451, 1974.

[14] S. W. Allen et al., Mon. Not. R. Astron. Soc. **346**, 593, 2003; H. V. Klapdor-Kleingrothause et al., Nucl. Instr. Methods Phys. Res., Sec. **A 522**, 371, 2004; G. Gratta et al., Int.J.Mod.Phys., **A19**, 1155, 2004.

[15] M. Tegmark, hep-ph/0503257.

## Appendix: The "strange" lepton flavor physics

In the QD-scenario, the lepton mass spectrum contains 4 mass-degenerate Majorana mass levels: three two-fold exactly-mass-degenerate CL mass levels $m_\tau$, $m_\mu$ and $m_e$ (three carrying charge Dirac states) plus one three-fold Quasi-Degenerate Majorana-neutrino mass level $m_\nu$. The three CL mass levels are highly hierarchical, with the hierarchy-rule approximately described by Eq.(27) in terms of powers of the constant $\alpha_o = e^{-5}$. Can this high CL mass-ratio hierarchy be extended so to include the fourth comparatively very low QD-neutrino mass level $m_\nu$? A probable answer, in the framework of lepton flavor mass-ratio phenomenology, see footnote [8], is a *factorial hierarchy,* which extends the sequence in (27):

$$m^2_{l+1}/m^2_l \cong \alpha_o^{\ell!}/2. \tag{A1}$$

The notations are $m_1=m_\tau$, $m_2=m_\mu$, $m_3=m_e$ and $m_4=m_\nu$. The factorial mass-ratio hierarchy between four lepton mass levels in (A1) is represented by three terms

$$m^2_\mu/m^2_\tau \cong \alpha_o/2, \ m^2_e/m^2_\mu \cong \alpha_o^2/2, \ m^2_\nu/m^2_e \cong \alpha_o^6/2. \tag{A2}$$

The absolute value of the QD-neutrino mass scale from (A2) is given by

$$m_\nu \cong \alpha_o^3 \ m_e/\sqrt{2} \cong 0.11 \text{ eV}. \tag{A3}$$

The change of the bare value $\alpha_o$ to the exact value of the low energy fine structure constant $\alpha$ could lead to an increased neutrino mass scale

$$m_\nu \cong \alpha^3 \ m_e/\sqrt{2} \cong 0.14 \text{ eV}. \tag{A4}$$

The estimations (A3) and (A4) are compatible with the cosmological neutrino bounds [15] for QD-neutrinos $m_\nu < 0.14$ eV (at 95%).

So, a drastically growing in the direction of smaller masses mass-ratio-hierarchy of the four lepton mass levels $m_\tau$, $m_\mu$, $m_e$ and $m_\nu$ may describe the extreme relative smallness of the QD-neutrino mass scale $m_\nu$ within the considered lepton mass-ratio phenomenology of the "strange", new lepton flavor physics beyond the electroweak theory.